\documentclass[apsrev4-2,prb,superscriptaddress,longbibliography,twocolumn]{revtex4-2}
\usepackage{color}
\usepackage{amsfonts,amsmath,amssymb}
\usepackage[dvipsnames]{xcolor}
\usepackage{tabularx,hhline,tikz,multirow,array}
\usepackage{bm,enumitem}
\usepackage{dcolumn,stmaryrd}
\usepackage{comment}
\usepackage{float} 
\usepackage[bookmarksnumbered,pdfpagelabels=true,plainpages=false,colorlinks=true,linkcolor=red,citecolor=red,urlcolor=red]{hyperref}

\allowdisplaybreaks

\begin{document}

\title{Probing Random-Bond Disorder Effects on Ferromagnetic Skyrmion Arrays}

\author{E. Iroulart}
\affiliation{Instituto de F\'isica de L\'iquidos y Sistemas Biol\'ogicos (IFLYSIB), UNLP-CONICET, Facultad de Ciencias Exactas, Universidad Nacional de La Plata, 1900 La Plata, Argentina}
\affiliation{Departamento de F\'isica, Facultad de Ciencias Exactas, Universidad Nacional de La Plata, 1900 La Plata, Argentina}

\author{F. A. G\'omez Albarrac\'in}
\affiliation{Instituto de F\'isica de L\'iquidos y Sistemas Biol\'ogicos (IFLYSIB), UNLP-CONICET, Facultad de Ciencias Exactas, Universidad Nacional de La Plata, 1900 La Plata, Argentina}
\affiliation{Departamento de Cs. B\'asicas, Facultad de Ingenier\'ia, Universidad Nacional de La Plata, 1900 La Plata, Argentina}

\author{H. D. Rosales}
\affiliation{Instituto de F\'isica de L\'iquidos y Sistemas Biol\'ogicos (IFLYSIB), UNLP-CONICET, Facultad de Ciencias Exactas, Universidad Nacional de La Plata, 1900 La Plata, Argentina}
\affiliation{Departamento de Cs. B\'asicas, Facultad de Ingenier\'ia, Universidad Nacional de La Plata,  1900 La Plata, Argentina}

\date{\today}

\begin{abstract}
In this work, we examined the impact of disorder in a ferromagnetic skyrmion  lattice by introducing random-bond disorder based on two different models for both exchange and Dzyaloshinskii-Moriya interactions: Model I, where both interactions present the same disorder distribution and thus the same local distortion, and Model II, where both interactions have different disorder distributions. Through extensive Monte Carlo simulations, we explored the effect of bond disorder on the emergent phases induced by an external magnetic field at different temperatures, varying the disorder amplitude across a range from weak (10\%) to strong (200\%) regimes. Our study shows that for both models, moderate disorder at low magnetic fields breaks the helical order and induces a topologically non-trivial bimeron phase. We also found that the skyrmion lattice phase loses its periodicity, but surprisingly, the resulting phases retain topological characteristics up to high disorder amplitudes. Moreover, a significant difference between both disorder models is found at large magnetic fields: while in the first model, the high-field skyrmion phase is suppressed by disorder, in the second model, this phase is enhanced and expanded in temperature and magnetic field. Furthermore, for the second model, intermediate values of disorder induce a diluted skyrmion phase and chiral textures emerging from the ferromagnetic phase at low temperatures. Our results are relevant for layered magnets and provide valuable insights into the intricate behavior of skyrmion systems under varying disorder conditions.
\end{abstract}

\maketitle
\section{Introduction}

Magnetic skyrmions, intriguing topological spin textures, hold significant promise for breakthroughs in energy-efficient data storage, low-power computing, and spintronic applications \cite{gobel2020beyond,zhang2023magnetic}. These swirling whirlpools of magnetization exhibit high stability and efficient manipulation, making them promising building blocks for future technologies \cite{finocchio2016magnetic,psaroudaki2021skyrmion,psaroudaki2023skyrmion,xia2023universal}. The conventional route to stabilizing skyrmions relies on the presence of the Dzyaloshinskii-Moriya (DM) interaction \cite{dzyaloshinsky1958thermodynamic,moriya1960anisotropic}, generally present in magnets without inversion symmetry, and the application of an external magnetic field along a threefold symmetry axis of different systems and materials \cite{bogdanov1994thermodynamically,roessler2006spontaneous,heinze2011spontaneous,nagaosa2013topological,rosales2015three,hayami2016bubble,hayami2022skyrmion,hayami2022multifarious}. 

In the quest for more exotic mechanisms to stabilize these topological textures, a wide variety of mechanisms have been shown to induce these textures, such as bond-dependent exchange anisotropy \cite{yi2009skyrmions,gao2020fractional,amoroso2020spontaneous,wang2021meron,rosales2022anisotropy}, the Ruderman-Kittel-Kasuya-Yosida (RKKY) interaction \cite{wang2020skyrmion}, higher-order exchange interactions \cite{paul2020role}, and magnetic frustration \cite{okubo2012multiple,mohylna2022spontaneous}. Recently, chiral spin liquids have emerged as a promising new avenue for stabilizing fluid phases of skyrmions in the presence of thermal fluctuations \cite{rosales2023skyrmion,gomez2024chiral}. In this regard, many studies have focused on the effects and consequences of disorder and lattice distortions in models hosting magnetic skyrmions\cite{diaz2017fluctuations,mirebeau2018spin,hoshino2018theory,chudnovsky2018skyrmion,reichhardt2022statics,liu2022disorder,henderson2022skyrmion,silva2014emergence,mohylna2023robustness,rosales2024robustness,bo2024suppression}. Considering disorder is crucial in real materials, which naturally exhibit crystal defects, distortions, or impurities, with disordered systems being ubiquitous in physics and materials science \cite{chaikin1995principles,sethna2021statistical}.
 
 A straightforward and extensively studied approach to introduce disorder in magnets is through the addition of 
 impurities inducing random local anisotropy,  missing spins, and holes, among other effects. Along this line, previous studies on the effect of impurities on skyrmion systems have revealed diverse effects on the stability and dynamics of skyrmions. For example, it has been shown that skyrmions can experience both weak and strong pinning effects depending on the sample thickness and material type, exhibiting a rich phenomenology of dynamics influenced by the nature of the disorder \cite{reichhardt2022statics}. Monte Carlo simulations of diluted 2D chiral magnets have demonstrated that nonmagnetic impurities can induce the formation of bimerons and distort the skyrmion lattice \cite{silva2014emergence}.
Furthermore, studies on frustrated Heisenberg antiferromagnets have shown that the skyrmion lattice phase is more robust against impurities than non-frustrated systems. Impurities distort the shape, size, and position of individual skyrmions, 
leading to a reduction of chirality and the eventual destabilization of the skyrmion lattice\cite{mohylna2023robustness}. These findings highlight that skyrmion systems can display diverse impurity-induced behaviors, such as different glassy states and phase transitions, which are essential for advancing their potential applications in technology.

Less explored is the effect of disorder through the Random-Bond-Disorder model (RBD) in two-dimensional magnets, particularly in skyrmion systems. Previous studies in non-skyrmion systems have shown intriguing results: in two-dimensional chiral antiferromagnets, infinitesimal RBD  shifts the ground state from noncollinearly ordered to a chirally ordered quasi-long-range-ordered state, transitioning to a short-range-ordered glassy state under stronger disorder \cite{dey2020random}. Similarly, investigations into the kagome antiferromagnet have shown that RBD leads to a spin liquid phase resembling jamming in granular media, which evolves into a conventional spin glass with increasing disorder strength\cite{bilitewski2017jammed}.

Motivated by the previous results on the effect of impurities and RBD in 2D magnets, our aim here is to explore how strong quenched random exchange interactions, arising from variations in bond angles caused by the chemical substitutions on nonmagnetic sites, affect the stability of skyrmions and non-topological phases (helical and fully polarized) that are typically found in skyrmion systems. To address these issues, we studied the effect of quenched RBD in a ferromagnetic square lattice model, combining exchange and in-plane Dzyaloshinskii-Moriya interactions. In particular, we considered two possible models for randomness and studied their effect in a typical ferromagnetic skyrmion system through extensive Monte Carlo simulations.

The rest of the paper is organized as follows. In Sec.~\ref{sec:Model}, we introduce the models and discuss the free-disorder phase diagram. In Sec.~\ref{sec:results}, we present our simulation analysis at finite temperature, identifying the topological (non-trivial) regions for the two models we have considered. Section~\ref{sec:conclusions} is devoted to the summary and conclusions.

\section{Model and Method}
\label{sec:Model}
We considered the classical ferromagnetic  Heisenberg model on a square lattice:

\begin{eqnarray}
H &=& -\sum_{\langle i,j \rangle}J_{ij} \vec{S}_i \cdot \vec{S}_j + \sum_{\langle i,j \rangle} \vec{D}_{ij} \cdot (\vec{S}_i \times \vec{S}_j)\nonumber \\
&&\quad - B\sum_i \vec{S}_i^z - A\sum_i (\vec{S}^z_i)^2 
\label{hamiltonian}
\end{eqnarray}

\noindent where the spin variables $\vec{S}_i$ are unimodular classical vectors at site $\vec{r}_i$, $J_{ij}$ is the exchange interaction, $\vec{D}_{ij}= D_{ij}\,(\vec{r}_j-\vec{r}_i)/\lvert\vec{r}_j-\vec{r}_i\rvert$ is the DM interaction—generally present in magnets without inversion symmetry—  with amplitude $D_{ij}$ between neighboring sites $i$ and $j$, the magnetic field $\vec{B}=(0,0,B)$ in the z-axis with strength $B$, $A$ is the out-of-plane magnetic anisotropy, and the sum $\sum_{\langle i,j \rangle}$ is taken over nearest-neighbors. It is worth pointing out the role of the magnetic anisotropy here. It plays an essential role in the magnetization process and the formation of magnetic skyrmions in ferromagnetic materials \cite{fattouhi2020formation,yuksel2021formation}. As the exchange and the DM interactions are already set, there is a delicate tradeoff between the Zeeman energy and the anisotropy; when the latter is weak, the system requires more Zeeman energy for emergence of skyrmions in the lattice.

To investigate the effect of random-bond disorder on the stability of skyrmions, we considered random variation for both the nearest-neighbor exchange constant $J_{ij}$ and the DM amplitude $D_{ij}$, as:
\begin{equation}
J_{ij} \xrightarrow{} J + \rho_{ij} \, \delta, \quad D_{ij} \xrightarrow{} D + \omega_{ij} \, \delta.
\label{eq:Jrdelta}
\end{equation}
\noindent where, for simplicity, we set up $J=1$ as the energy scale, we chose $D=1$ (which is large enough to stabilize skyrmions) and we fixed the anisotropy to $A=0.1$ throughout the rest of this work. In  Eqs.~(\ref{eq:Jrdelta}), the parameters $\rho_{ij}$ and $\omega_{ij}$ represent random numbers for each lattice bond chosen uniformly within $[-1,1]$. We systematically varied the disorder amplitude $\delta$, ranging from 0 to a maximum of $\delta=2$, corresponding to a possible variation of up to 200$\%$ in the interactions. We examined two specific scenarios, which we will refer to as Model I and Model II in the rest of this work: 

\begin{itemize}
\item{\textbf{Model I:} outlined in subsection \ref{sec:same}, where  $\rho_{ij}$ equals $\omega_{ij}$, meaning that both interactions have the same disorder distribution (i.e., the $D_{ij}/J_{ij}$ ratio is fixed to $D_{ij}/J_{ij}=D/J = 1$).}  
\item{\textbf{Model II:} discussed in subsection \ref{sec:dif}, where $\rho_{ij}$ and $\omega_{ij}$ differ, implying that the disorder in both interactions can differ considerably between them (i.e., the $D_{ij}/J_{ij}$ ratio changes in each bond).}

\end{itemize}

\begin{figure}[htb] 
\centering
\includegraphics[width=0.8\columnwidth]{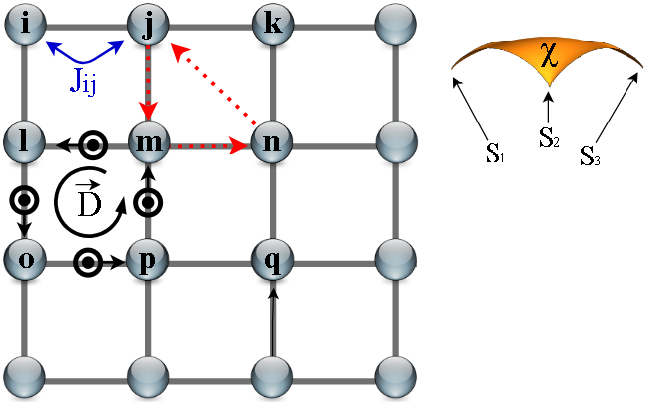}
\caption{(Left) Illustration of the square lattice, indicating nearest neighbors exchange  $J_{ij}$ between sites $(i,j)$ (color blue) and Dzyaloshinskii-Moriya interaction $D_{ij}$ between pairs of sites $(m,l),(l,o),(o,p),(p,m)$(color black). Red dashed arrows indicate the direction that three spins at sites $(m,n,l)$ are taken to calculate the scalar chirality  (Eq.\ref{eq:chir}).
(right) Three spin vectors $S_1$, $S_2$, and $S_3$ form a solid angle, illustrating the geometric origin of the scalar chirality $\chi$ in spin systems.
}
\label{fig:lattice}
\end{figure}
\subsection{Monte Carlo Simulations}
\label{sec:MCsimulations}

We performed Monte Carlo simulations based on the standard Metropolis algorithm combined with the over-relaxation method \cite{brown1987overrelaxed,creutz1987overrelaxation}. We implemented periodic boundary conditions for a square lattice with $N=L^2$ sites, with $L=30,48,60$. We gradually cooled down the system exponentially from high temperature down to $T=8.7\times 10^{-3}$, taking $10^5$ Monte Carlo steps for initialization and four times as much for measurements. Averages were taken over 100 independent disorder realizations. To characterize the different phases, we computed the scalar chirality $\chi$ to identify various types of chiral states, defined as
\begin{equation}
    \chi = \frac{1}{8\pi}  \left\langle \sum_{m}\vec{S}_{m} \cdot \left( \vec{S}_{n} \times \vec{S}_{j} \right) \right\rangle,
    \label{eq:chir}
\end{equation}
where $\{\vec{S}_{m},\vec{S}_{n},\vec{S}_{j}\}$ are three neighboring spins, as depicted in Fig.~\ref{fig:lattice}. The quantity $\chi$ is the discrete version of the topological charge for skyrmions. Accordingly, the skyrmion phase is characterized by a non-zero $\chi\neq 0$, while the helical and fully polarized (ferromagnetic) phases present  $\chi=0$.
Finally, we computed the structure factor $S_{\perp}(\vec{q})$ at low temperatures, which has distinct characteristics for the helical, skyrmion crystal, and fully polarized phases
\begin{equation}
S_{\perp}(\vec{q}) = \frac{1}{N} \left\langle \left| \sum_i \vec{S}^x_{i} e^{\vec{q}\cdot \vec{r}_i} \right|^2 + \left| \sum_i \vec{S}^y_{i} e^{\vec{q}\cdot \vec{r}_i} \right|^2 \right\rangle
    \label{sfxy}
\end{equation}
We utilized the structure factor to further explore the different phases in the system. A skyrmion crystal phase may be understood as a superposition of three single-$q$ state helices \cite{muhlbauer2009skyrmion,jonietz2010spin,munzer2010skyrmion,yu2010real}, resulting in a triple-$q$ pattern (fulfilling the condition $\sum_i \vec{q}_{i}=0$), reflected in six bright peaks in the structure factor, comparable with neutron scattering measurements. 

\begin{figure*}[htb!]
\includegraphics[width=0.9\textwidth]{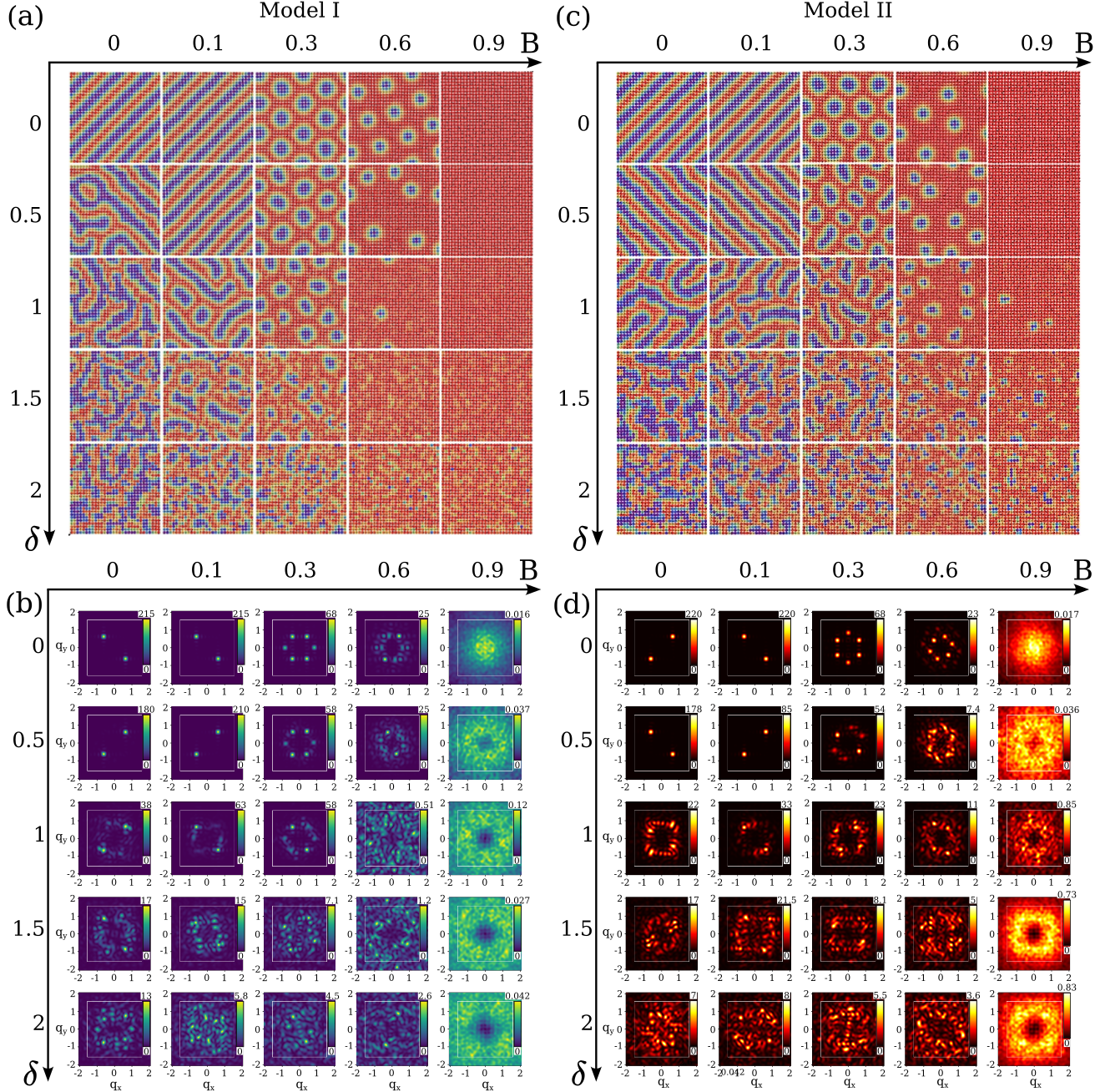}
\caption{Representative magnetic configurations (panels a,c) and structure factors (panels b,d)  for several disorder amplitudes $\delta$ and magnetic fields $B$ at the lowest simulated temperatures $T=8.7\times10^{-3}$ for  Model I (left column) and Model  II (right column).}
\label{fig:figtodo}
\end{figure*}

Before presenting the results for $\delta>0$, let us briefly review the situation with zero disorder $\delta=0$. The magnetic phase diagram for the model defined by Eq.~(\ref{hamiltonian}) without disorder (i.e., $\delta=0$ in Eq.~(\ref{eq:Jrdelta})) has been intensively discussed in many previous works \cite{bogdanov1989thermodynamically,roessler2006spontaneous,yu2010real,ezawa2011compact}. It is well established that at low temperatures and moderate DM $D_{ij}=D$, for $D/J=1$, the low temperature phase diagram consists of three well-defined phases: the single-q helical (HL) phase for $0 \le B < B_{c1} \simeq 0.3$; a periodic array of magnetic skyrmions, the skyrmion lattice (SkX) phase for $B_{c1} \leq B \leq B_{c2} \simeq 0.6$ and the fully polarized (FP) phase for $B_{c2} < B$, with representative snapshots shown in Fig.~\ref{fig:figtodo}(a,b). At finite temperature, thermal fluctuations induce two types of intermediate phases \cite{ezawa2011compact}: a bimeron (Bm) phase and a skyrmion gas (SkG). The Bm phase emerges as part of the magnetisation process as the magnetic field is increased from the helical phase: the helices are broken, and thus bimerons or ``elongated skyrmions'' are formed \cite{leonov2024}. Notice that this definition of bimeron differs from that where a bimeron is formed by two merons \cite{Gobel2019}. As for the SkG, it is an intermediate texture between the skyrmion lattice and the fully polarized phase, where the skyrmions do not form a periodic arrangement.

The following section will be dedicated to describing the effect of disorder through the disorder amplitude $\delta$, following the two models described above.

%

\section{Results}
\label{sec:results}

In this section, we thoroughly investigate the proposed model in Eq.~(\ref{hamiltonian}) by examining disorder through linear random changes in the strength of the couplings ($J_{ij}, D_{ij}$),  as described in Eq.~(\ref{eq:Jrdelta}). 
In this regard, we use extensive Monte Carlo simulations to explore these two proposals for bond disorder across $0 \leq \delta \leq 2$ and analyze their effects on the skyrmion phases under various temperatures and magnetic fields. It is well established that random bond disorder is bound to induce glassy behavior, as has been shown even in highly degenerate models \cite{shinaoka2014effect,bilitewski2017jammed,dey2020random}. Therefore, the emergent phases in this work are expected to have significantly slower dynamics than in a no-disorder case. Since the main goal of this work is to explore the effect of disorder in the emergence of topological textures, we defer the exploration of the possible glassy dynamics for future work.
\subsection{Disorder Model I: $\rho_{ij}=w_{ij}$}
\label{sec:same}

In Fig.~\ref{fig:figtodo} we show the evolution of representative snapshots (panel (a)) and structure factors (panel (c)) for different values of the disorder amplitude $\delta$ for various magnetic fields $B$ at the lowest simulated temperature ($T=8.7\times10^{-3}$). In the no-disorder case ($\delta=0$, first row in panel a)), as the magnetic field increases, the system transitions from the helical phase ($B=0, 0.1$) to the skyrmion crystal phase ($B=0.3$), going through a skyrmion gas phase ($B=0.6$) before ultimately reaching the fully polarized phase ($B=0.9$). The effect of including disorder on these different phases ($\delta\neq 0$) can be observed along the different columns in Figs.~\ref{fig:figtodo} (a) and (c), as described below:

\begin{itemize}

\item{\textit{Helical phase:} In the case of the helical phase, one might initially think that the effect of disorder would be solely to distort the helices at low $\delta$ values and eventually disintegrate them at high $\delta$ values. However, interestingly, we observe that increasing disorder induces the fragmentation of the helical stripes into skyrmions and bimerons (which have the same topological number as a skyrmion \cite{ezawa2011compact,silva2014emergence,rosales2023skyrmion,gomez2024chiral}). In Fig.~\ref{fig:figtodo}(a), panel ($\delta=1, B=0.1$), we show a representative snapshot of skyrmions and bimerons embedded in a ``labyrinth''-domain structure. Previous works have shown that bimerons can also be stabilized by vacancies randomly distributed over the system. In those studies, the mechanism of bimeron stabilization is induced by an effective local magnetic field introduced by spin vacancies \cite{silva2014emergence}. Here, we show that random bonds can create bimerons+skyrmions in the helical phase (in an effect similar to that of thermal fluctuations), but only for $\delta<=1$, i.e., for values of the disorder amplitude that imply that the exchange interaction remains ferromagnetic, $0\leq J_{ij}\leq J$, and the DM interaction has always the same orientation $0\leq D_{ij}\leq J$. For $\delta>1$,  $J_{ij}$ could be negative, meaning that the model includes both ferromagnetic and antiferromagnetic couplings (and DM interaction with different orientations), and thus local magnetic frustration emerges. In this cases, the magnetic configurations becomes more irregular and eventually disintegrate for larger $\delta$.
}
\item{\textit{Skyrmion phase:} In the skyrmion lattice phase, low disorder values distort the skyrmion profiles, breaking their periodic arrangement (Fig.~\ref{fig:figtodo}(a), column $B=0.3$). Interestingly, unlike the inclusion of local impurities—which decreases the number of skyrmions and leads to the formation of bimerons \cite{silva2014emergence}-in the skyrmion lattice phase, the skyrmions are robust and the number of skyrmions remains constant. This constancy is observed up to $\delta \leq 1$, similar to what occurs in the helical phase. For $\delta > 1$, skyrmions become more irregular and eventually disintegrate at higher $\delta$ values.

Conversely, in the skyrmion gas phase ($B=0.6$), even at low disorder values, the number of skyrmions sharply decreases until $\delta = 1$, suggesting that the skyrmion gas is less robust than the SkX, and more susceptible to local coupling changes. 
Beyond this point, magnetic frustration arises due to $J_{ij}$ taking negative values and the possible change of direction in the DM interaction, and a disordered texture emerges.
}
\item{\textit{Fully polarized phase:} Finally, the fully polarized phase remains stable up to $\delta = 1$. However, for $\delta > 1$, as described above, local antiferromagnetic arrangements emerge due to the sign change in $J_{ij}$, and thus the magnetisation drops.
}
\end{itemize}

To explore the consequences of disorder in reciprocal space, we calculated the perpendicular component of the structure factor ($S_\perp(\vec{q})$) at low temperatures as outlined in Eq.~(\ref{sfxy}). The results are depicted in Fig.~\ref{fig:figtodo}(c), where we observe the evolution of $S_\perp(\vec{q})$ as we systematically increase the disorder amplitude for the helical phase (with $B=0.1$), the skyrmion phase (with ($B=0.3$), and the ferromagnetic phase (with $B=0.9$). Notably, as the disorder amplitude grows, an intriguing transformation occurs. The distinct Bragg peaks, initially sharp and well-defined, gradually blur and diffuse, especially when approaching the highest disorder amplitude values ($\delta > 1 $). Thus, we see that the trend towards a disordered lattice structure is reflected in reciprocal space, reinforcing the correlation between disorder amplitude and the magnetic behavior of the system.

\begin{figure}[H]
\includegraphics[width=1.0\columnwidth]{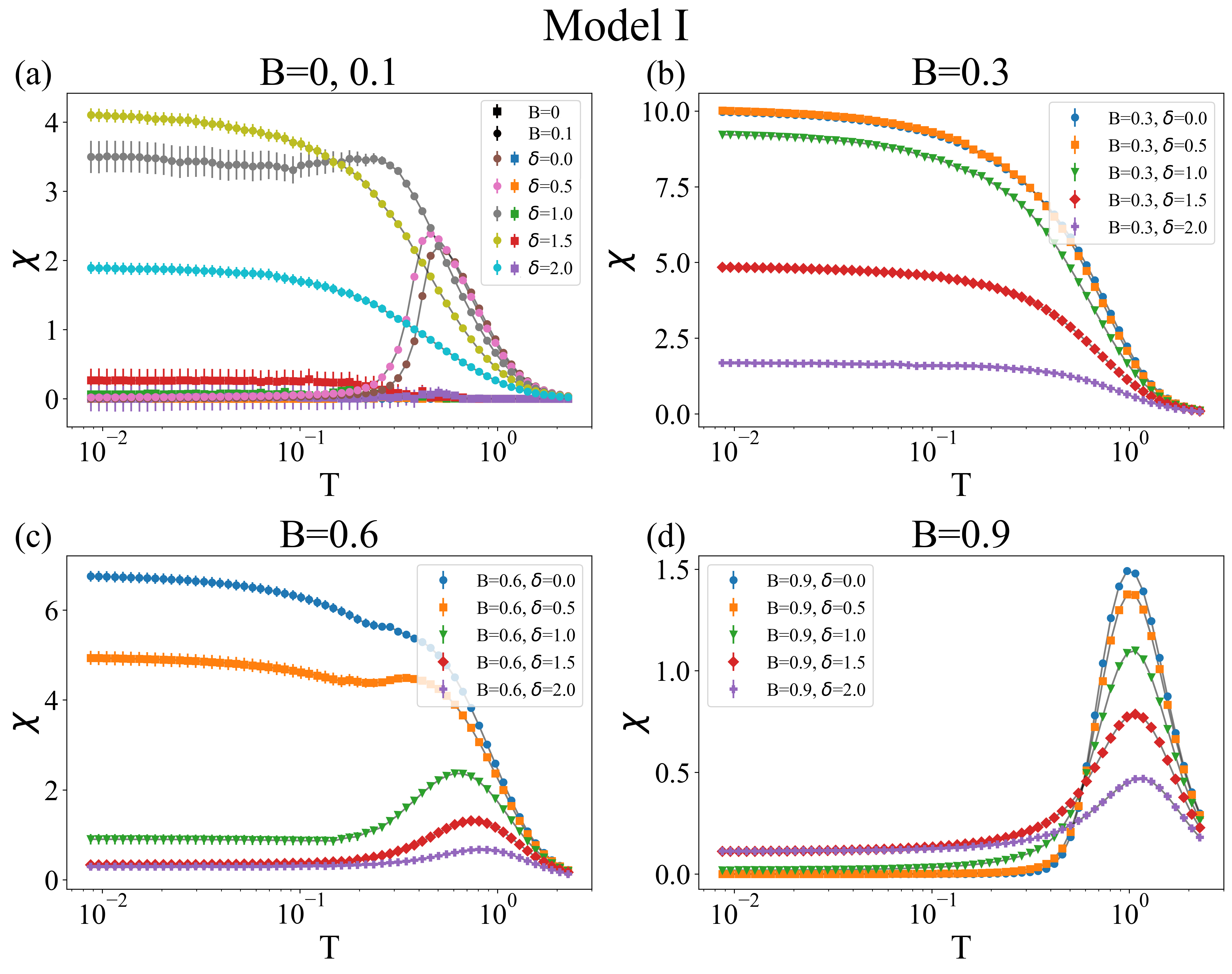}
\caption{Scalar chirality as a function of temperature for a set of disorder amplitudes $\delta =\{0, 0.5, 1.0, 1.5, 2.0\}$ in the first disorder model: (a) $B=0.0$ and $B=0.1$, helical phase, (b) $B=0.3$ skyrmion crystal phase,  (c) $B=0.6$ skyrmion gas phase, and (c) $B=0.9$  ferromagnetic or fully polarized phase.} 
\label{Chir_vs_T_delta_m1}
\end{figure}

Since the topological phases are characterized by the scalar chirality $\chi$, we study how the effect of disorder is reflected in this order parameter. In Fig.~\ref{Chir_vs_T_delta_m1}, we plot ﬁeld and temperature dependencies of the chirality for selected values of the disorder amplitudes $\delta = \{0, 0.5, 1.0, 1.5, 2.0\}$. The data is presented for four representative cases: (a) the helical phase with $B=0$  and $B=0.1$ ; (b) the skyrmion lattice at $B=0.3$; (c) the skyrmion gas at $B=0.6$; and (d) the fully polarized (FP) phase at $B=0.9$.  The average is taken over 100 independent realizations, and the error bars are calculated as the statistical error.

Firstly, in the helical state for $B=0$ (Fig.~\ref{Chir_vs_T_delta_m1}(a)), $\chi$ is almost zero for all disorder amplitudes $\delta$. However, under a small magnetic field ($B=0.1$), the situation changes significantly: for low disorder amplitudes ($\delta <0.5$),  $\chi$ is approximately zero at low temperature, but at intermediate temperatures, thermal fluctuations stabilize bimerons, whose entropy is much larger than that of the helix\cite{ezawa2011compact}.  This increase in chirality appears within a well-defined temperature window. Interestingly, for $\delta=1$, the region of bimerons (now including skyrmions textures) extends down to the lowest temperature. Finally, as $\delta$ is further increased, causing some bonds to switch from ferromagnetic to antiferromagnetic ($ \delta > 1$), the low-temperature values of $\chi$ decrease, the characteristic scale for helices and skyrmions disappears, and the resulting textures become distorted, as shown in the bottom panels of Fig.~\ref{fig:figtodo}(a).

 In the skyrmion crystal, skyrmion gas, and in the field-polarized phases (Fig.~\ref{Chir_vs_T_delta_m1}(b), (c) and (d)), disorder induces a distortion of the skyrmion profiles, consequently reducing the magnitude of the chirality. In the skyrmion crystal case ($B=0.3$), panel (c)), $\chi$ decreases for higher $\delta$ as the skyrmion crystal transitions into a possibly glassy phase, and the skyrmions themselves lose shape under stronger disorder. Notably, even at $\delta \sim 2 $, the chirality only decreases to about $20\%$ of its value with no disorder, remaining non-zero.  
 
 For higher fields, the situation is more straightforward. In the skyrmion gas case ($B=0.6$), panel (c)),  the chirality drops faster than for the skyrmion crystal case, as the number of skyrmions quickly decreases with disorder, eventually reaching zero. This effect of skyrmion distortion and reduction in chirality is similar to what is observed in the presence of non-magnetic impurities \cite{mohylna2023robustness}. Finally, in the fully polarized phase at ($B=0.9$), the peak in $\chi$ at intermediate temperatures is a well-known feature of the system without disorder, indicating a skyrmion gas at intermediate temperatures\cite{ezawa2011compact}. This feature is washed out by disorder. As $\delta$ increases, disorder drives the system away from the fully polarized phase, reducing the magnetization (see Fig.~\ref{fig:figtodo}(a)).

\begin{figure} 
\centering
\includegraphics[width=1.0\columnwidth]{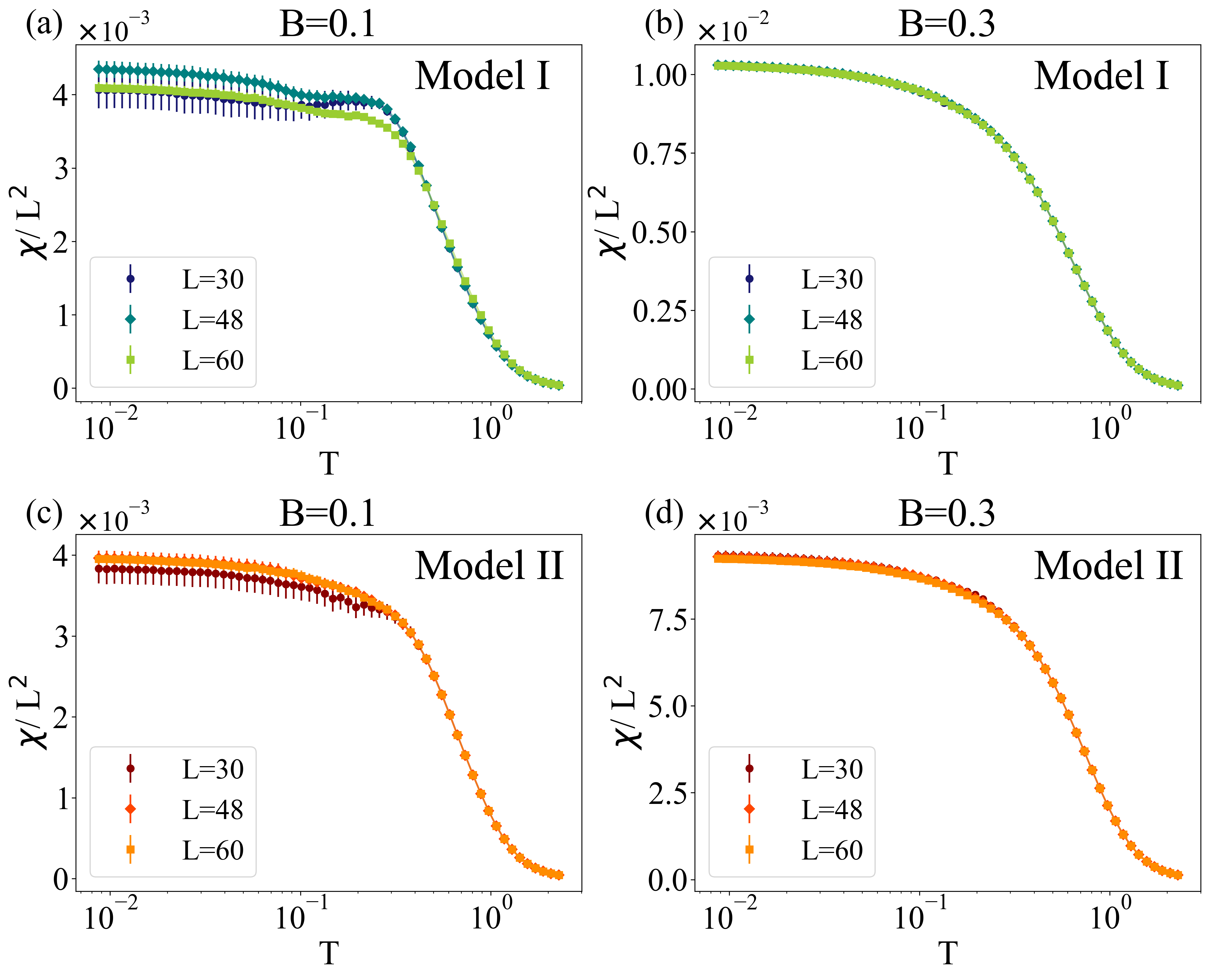}
\caption{Chirality density ($\chi/L^2$) as a function of temperature for a fixed disorder amplitude $\delta=1.0$ for lattice sizes $L=30, 48, 60$. Panels (a) and (b) show the helical and skyrmion phases, respectively, in Model I, while panels (c) and (d) showcase the corresponding phases utilized for Model II. }
\label{ChirL2_vs_T_SIZE}
\end{figure}
To validate our findings concerning system size, in Fig.~\ref{ChirL2_vs_T_SIZE} we show the temperature dependencies of the chirality scaled by system size, $\chi/L^2$ (chirality density), for two separate phases: helical and skyrmion. This is done for a fixed $\delta=1.0$ with panels (a, b) showing results for Model I and panels (c, d) for Model II, for three different lattice sizes $L=30, 48, 60$.

\begin{figure}[H] 
\centering
\includegraphics[width=1.0\columnwidth]{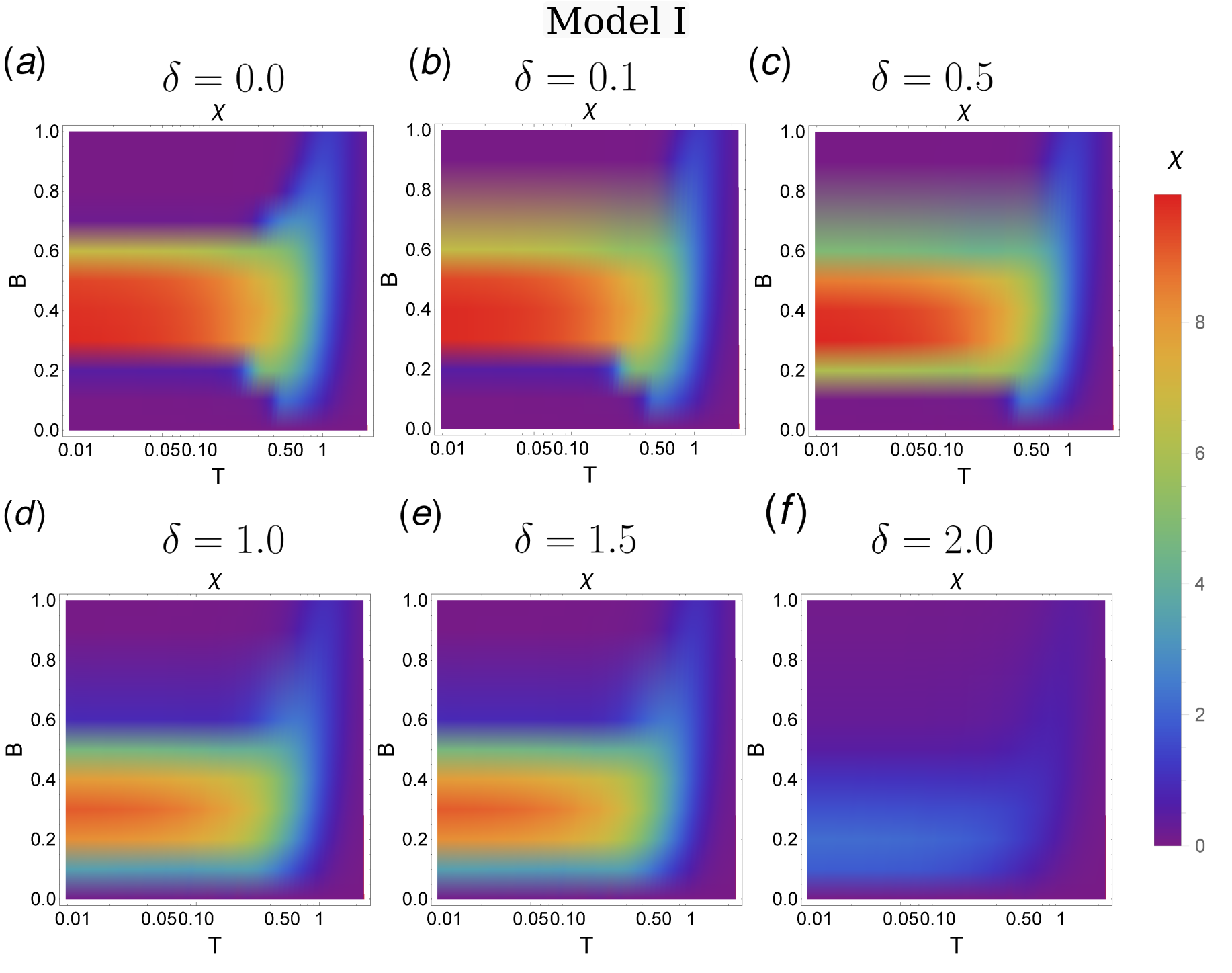}
\caption{Evolution of the density plot of the scalar chirality $\chi$  in the $B-T$  phase space  for several disorder amplitudes $\delta=\{0.0, 0.1, 0.5, 1.0, 1.5, 2.0\}$ for Model I. Chirality is scaled to the highest value, that at $\delta=0$ in the skyrmion crystal phase at low temperature.} 
\label{DP_6_dJ_M1}
\end{figure}

All the previous analyses are summarized in the phase diagrams shown in Fig.~\ref{DP_6_dJ_M1}, where we plot chirality $\chi$ as a function of magnetic field  $B$ and temperature $T$ for various disorder amplitudes $\delta$. For a better comparison, the chirality is normalized to its maximum value at $\delta=0$ in the skyrmion crystal phase, at the lowest temperature. As the disorder is introduced, the region of maximum chirality shifts toward lower magnetic fields until, for high disorder, the formation of skyrmion-like textures is prevented. The most remarkable feature is that in the helical region for $\delta = 0$, moderate values of disorder ($\delta<1$) induce the appearance of topologically non-trivial arrangements of bimerons at lower magnetic fields and temperatures. One significant feature of this model is that at higher fields, the chirality in the skyrmion gas phase is abruptly suppressed by disorder, persisting only for $\delta < 0.5$.
%

\subsection{Disorder Model II: $\rho_{ij}\neq w_{ij}$}
\label{sec:dif}

\begin{figure}[htb]
\includegraphics[width=1.0\columnwidth]{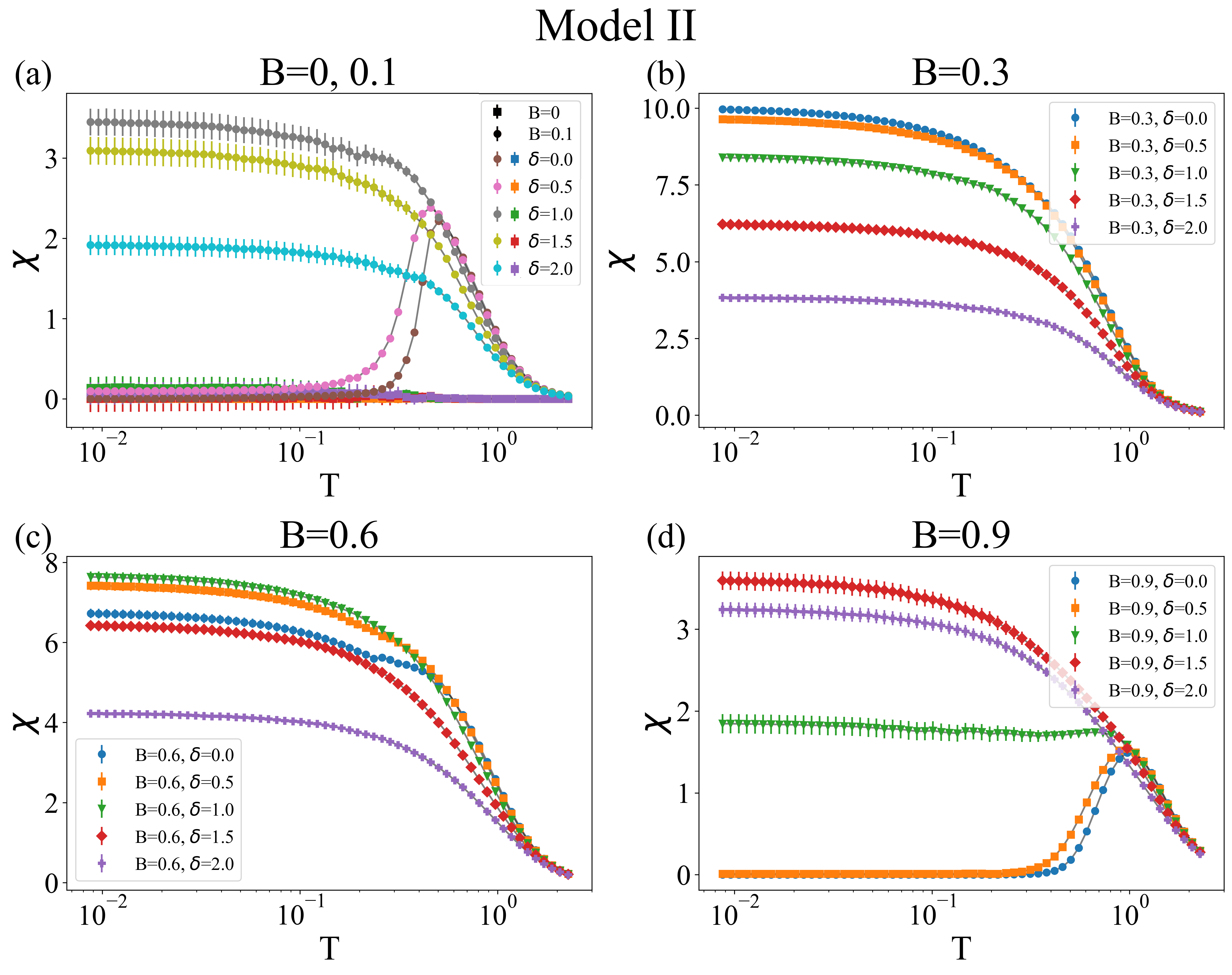}
\caption{Scalar chirality as a function of temperature (in logarithmic scale) for a set of disorder amplitudes $\delta =\{0, 0.5, 1.0, 1.5, 2.0\}$ in the second disorder model: (a) $B=0.0$ and $B=0.1$, helical phase, (b) $B=0.3$ skyrmion crystal phase,  (c) $B=0.6$ skyrmion gas phase, and (c) $B=0.9$  ferromagnetic or fully polarized phase.} 
\label{Chir_vs_T_delta_m2}
\end{figure}

We now turn to our second proposal to introduce bond disorder. Here, we also defined a disorder amplitude $\delta$ for both the exchange and DM magnitudes, but we sort independent random numbers $\{\rho_{ij},w_{ij}\}$ in the expression of the interactions defined in Eq.~(\ref{eq:Jrdelta}). To enable better comparison with Model I from the previous section (\ref{sec:same}), Fig.~\ref{fig:figtodo} (panels (b) and (d)) shows representative snapshots and structure factors, illustrating the system's evolution for different disorder amplitudes $\delta>0$ and magnetic field strengths $B$ at the same lowest simulated temperature ($T=8.7\times10^{-3}$). In Fig.~\ref{Chir_vs_T_delta_m2}, as for Model I, we plot $\chi$ as a function of temperature for the chosen values of disorder amplitude and magnetic fields, and we validate our results with system size for the helical ($B=0.1$) and skyrmion lattice phase ($B=0.3)$ in Fig.~\ref{ChirL2_vs_T_SIZE}, panels (c) and (d).  We describe these results below: 

\begin{itemize}
    \item{\textit{Helical phase:} In Fig.~\ref{fig:figtodo} we compare the principal results of both models: Model I in panels (a) and (c) and Model II in panels (b) and (d). In the helical phases ($B=0$ and $0.1$) both models present a similar behavior: increasing disorder induces the fragmentation of the helical stripes into skyrmions and bimerons. In particular, at $\delta=1$ and $B=0.1$ we observe that skyrmions and bimerons emerge in the ``labyrinth''- like domain structure. Finally, as $\delta$ is further increased, some bonds switch from ferromagnetic to antiferromagnetic ($ \delta > 1$) and the characteristic scale for helices disappear, and the resulting textures become nearly fully disordered. 
 }
\item{\textit{Skyrmion phase:} In the case of the skyrmion lattice phase (shown here for $B=0.3$), we found that, in both models, low disorder values distort the skyrmion pattern, breaking their periodic arrangement. The first main difference between disorder models arises in the skyrmion gas phase ($B=0.6$). Unlike Model I, Model II shows an increase in the number of skyrmions as $\delta$ increases, which reduces the effective skyrmion size. As shown in Figs.~\ref{Chir_vs_T_delta_m1} and \ref{Chir_vs_T_delta_m2}, while in Model I, the chirality decreases with increasing $\delta$ (Fig.~\ref{Chir_vs_T_delta_m1}(c)), in Model II, the chirality (and the number of skyrmions) initially increases up to a maximum value at $\delta = 1$, and then decreases, eventually leading to a disordered magnetic phase (see Fig.~\ref{Chir_vs_T_delta_m2}(c)). The distinct randomness in $J_{ij}$ and $D_{ij}$ modifies the local $D/J$ ratio, promoting skyrmion topological states. For $\delta > 1$, $J_{ij}$ and $D_{ij}$ can change signs, resulting in magnetic frustration and a disordered magnetic state.}
\item{\textit{Fully polarized phase:} Examining Figs.~\ref{fig:figtodo}, \ref{Chir_vs_T_delta_m1} and \ref{Chir_vs_T_delta_m2} and comparing both models (Model I and II), it is evident that the main difference arises in the fully polarized (FP) phase. While in Model I, increasing $\delta$ generally leads the system toward a magnetically disordered phase, in Model II, the randomness in the $D_{ij}/J_{ij}$ values induces the appearance of isolated skyrmions. This persists for $\delta \leq 1$, but for $\delta = 1.5$ and $2$, the skyrmions are deformed and get smaller, resulting in a magnetically disordered chiral phase (see Fig.~\ref{Chir_vs_T_delta_m2} (d)).}
\end{itemize}

Overall, compared to Model I, focusing in Model II at higher fields, we find a significant change for larger disorder. Interestingly,  in this case, we note that the skyrmion gas phase persists with disorder, and, most interestingly, we also note the tendency of the disorder to form a skyrmion gas from the FP phase, as seen in Fig.~\ref{fig:figtodo}(c,d). For $\delta \sim 1$, we see that skyrmion-like textures appear in the FP background. Larger values of disorder imply possible changes in the sign for the exchange and  DM interactions, giving rise to a stronger frustration and thus a noticeable change in the scale of the textures, eventually destroying the skyrmions. We contrast the effects of disorder for both models in subsection   \ref{subsec:comp}.

\begin{figure}[H]
\centering
\includegraphics[width=1.0\columnwidth]{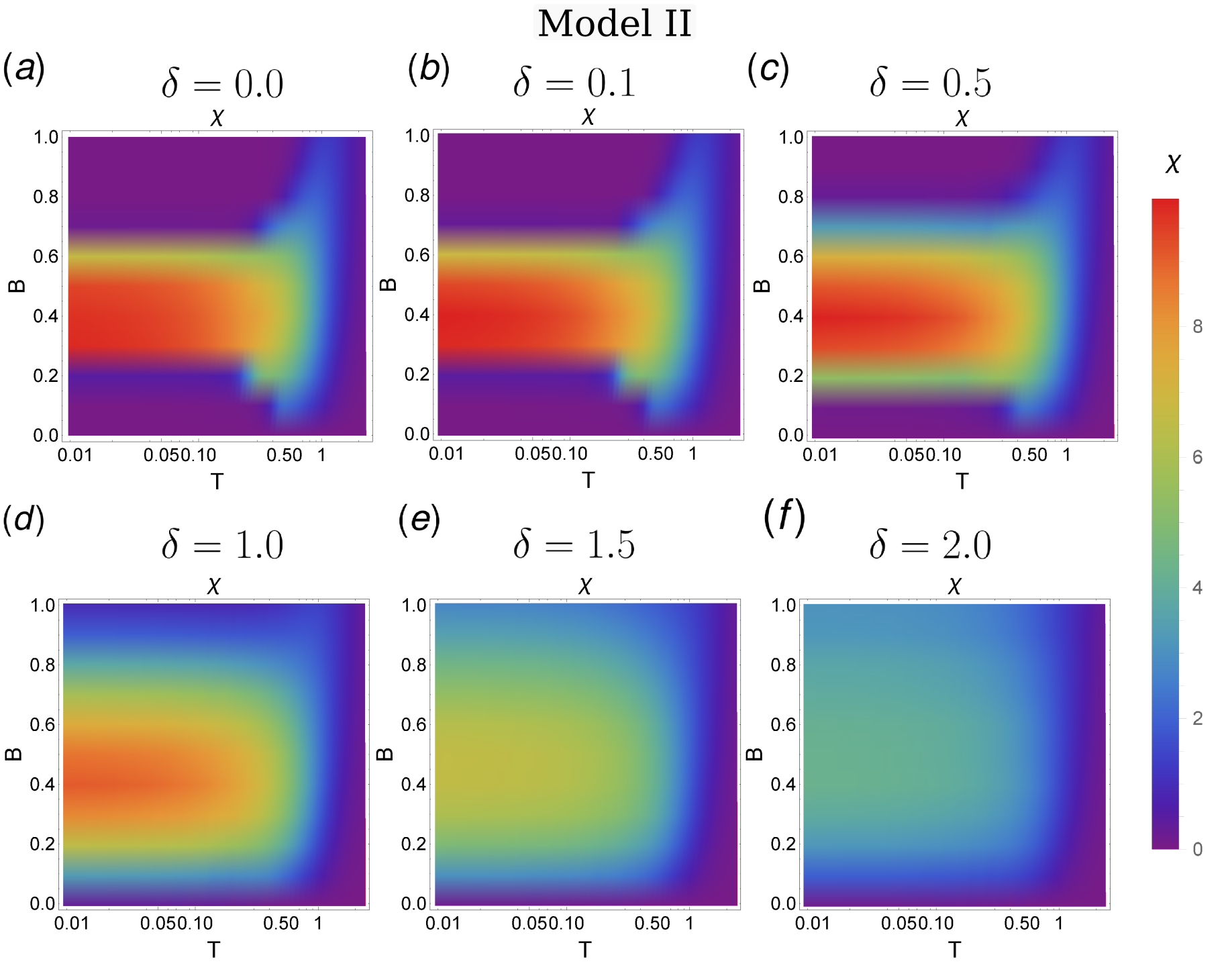}
\caption{Evolution of the density plot of the scalar chirality $\chi$  in the $B-T$ (in logarithmic temperature scale) phase space  for several disorder amplitudes $\delta=\{0.0, 0.1, 0.5, 1.0, 1.5, 2.0\}$ in Model II.  Chirality is scaled to
the highest value, that at $\delta=0$  in the skyrmion crystal phase.}
\label{DP_6_dJ}
\end{figure}

Additionally, to delve into the effect of randomness in the stabilization of topological textures with temperature in  Model II, as in the case for Model I, we plot the chirality $\chi$ as a function of magnetic field ($B$) and temperature ($T$) for several disorder amplitudes $\delta$ in Fig.~\ref{DP_6_dJ}, normalized to the highest value in the skyrmion lattice phase for $\delta=0$. We find that, contrary to what occurs in Model I, the chiral region expands with increasing disorder, occupying a larger portion of the phase diagram at both higher and lower magnetic fields. Additionally, we observe that at high disorder ($\delta>1$), even though the chirality is significantly lower than in the typical skyrmion phase diagram, it remains higher for intermediate fields and across a broad range of temperatures compared to the first bond disorder model (see Fig.~\ref{Chir_vs_T_delta_m2} (c) and (d)).

\begin{figure*}[ht!]
\includegraphics[width=0.8\textwidth]{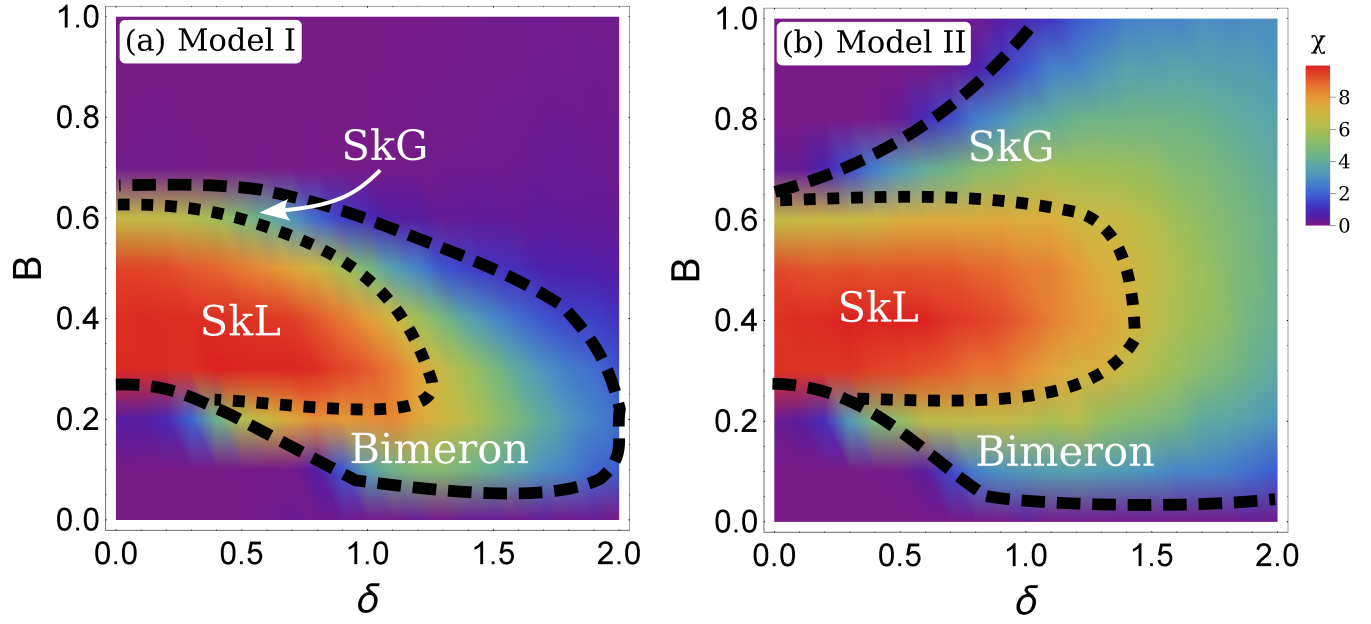}
\caption{(top) Density plots of scalar chirality $\chi$ in the $B-\delta$ phase space at low temperature ($T=8.7\times10^{-3}$) for  Model I (panel a) and Model II (panel b). The dashed lines indicating the boundaries of the phases were terminated using $\chi$.} 
\label{fig:DP_last_T_bothmodels}
\end{figure*}
\subsection{Comparison between disorder models \label{subsec:comp}}
After studying disorder in the emergent topological textures of a well-known skyrmion system through two RBD models, we focus here on the similarities and differences that arise between the effects of these two RBD proposals, Model I and Model II, described in Sec.~\ref{sec:Model}.  The distinction in the definition of Model I and Model II is simple: in Model I, the $D_{ij}/J_{ij}$ ratio is fixed, whereas in Model II the $D_{ij}/J_{ij}$ ratio is random in each bond. Surprisingly, as we discuss in detail below, although simple, this variation between models amounts to clear different outcomes in the magnetic system.

Consequently, to summarize our results and provide a clearer comparison between the two RBD models we have studied, in Fig.~\ref{fig:DP_last_T_bothmodels}, we present side-by-side the $B-\delta$ phase diagrams constructed using the chirality at the lowest temperature $T=8.7\times 10^{-3}$. Results for Models I and II are presented in panels (a) and (b), respectively. 

In both models, a noticeable feature is that the skyrmion crystal phase, at intermediate fields ($B\sim 0.3, 0.4$), retains its chirality even for higher disorders. At lower fields,  chiral phases are stabilized, due to disorder ``breaking'' the helical structure and forming bimerons. We also see that the most significant changes in both phase diagrams occur for higher amplitudes of disorder, $\delta>1$, where both the exchange coupling $J_{ij}$ and the DM amplitude $D_{ij}$ may change sign (see Eq.~(\ref{eq:Jrdelta})). A random change of sign in $J_{ij}$ in some bonds implies the emergence of local magnetic frustration due to the competition between ferromagnetic and antiferromagnetic interactions, which affects the scale of the textures (width of helices, skyrmion size), as shown in the snapshots in Fig.~\ref{fig:figtodo}. 

As discussed in the previous subsection, significant differences between models are seen as the magnetic field is increased, beyond the skyrmion lattice phase. For Model I, we see that the effect of moderate disorder is to shift the chiral region only at lower magnetic fields. At higher fields, topological phases such as the skyrmion gas are more susceptible and disintegrate due to disorder, leading to an abrupt suppression of chirality. On the other hand, for Model II, disorder stabilizes skyrmions and topological textures, ``expanding'' the region of chirality throughout the los temperature $B-\delta$ phase diagram, particularly at higher magnetic fields, where we find that disorder induces the emergence of skyrmion textures in the field polarized phase. The distinct behaviour of the emergent low-temperature phases at higher fields between both disorder models is also reflected in reciprocal space. Comparison of the structure factors depicted in Fig.~\ref{fig:figtodo}, panels (b) and (d), shows that for Model II a ring-like pattern associated with quenched skyrmion gas textures holds for $\delta \lesssim 1$ for $B=0.6,0.9$. At lower fields, similar patterns are found for both models.

The key to these variances between models seems to lie in the fact that for Model I, the $D_{ij}/J_{ij}$ ratio is fixed, and thus the effective model in each interaction may be thought as the same ``exchange + DM'' model at different magnetic fields. Thus, for example for $\delta = 1$, for a uniform distribution of randomness, half of the interactions are equal to a model with $J_{ij}=D_{ij}$ and higher magnetic fields (since $J_{ij}=D_{ij}<1$) and the other half to a model with lower magnetic fields ($J_{ij}=D_{ij}>1$). Therefore, in cases like the skyrmion gas phase, there is a competition where roughly half of the interactions favour a higher field phase, i. e. the field polarized phase, and the skyrmions are suppressed. In contrast, in Model II the strength of the exchange and DM interactions is not equal, and, recalling that the $D_{ij}/J_{ij}$ ratio is related to the skyrmion size (a smaller ratio implies larger skyrmion radii and lower saturation values), in each bond, a different scale of textures and effective magnetic fields is favoured. At higher external fields $B$, there may be effective bond models for smaller skyrmions at larger fields, and thus for large disorder amplitude $\delta$ this competition induces the stabilization of skyrmion textures. 

%
\section{Discussion and Conclusions}
\label{sec:conclusions}

The mail goal of this work is to study the effect of random disorder in the emergence of topological textures in a typical skyrmion model. To this end, we analyzed the effect of disorder in a ferromagnetic skyrmion square lattice with exchange and DM interactions by introducing random-bond disorder model, varying the disorder amplitude from 10\% to 200\% of the interactions and using two proposals: one in which we sort the same random number for the exchange and DM interactions (Model I), and another model that consists of sorting two different random numbers for each interaction (Model II). Thus, the $D_{ij}/J_{ij}$ ratio is fixed in each bond in Model I, and changes randomly in Model II.

 We found that, for both disorder models, at low magnetic fields ($B< 0.2$), disorder induces fragmentation of helical stripes into bimerons, leading to the emergence of topologically non-trivial phases. Additionally, intermediate disorder values break the periodicity of the skyrmion lattice found at intermediate fields, resulting in an amorphous skyrmion lattice. This effect becomes particularly pronounced for high disorder amplitudes, ultimately leading the system into a magnetically disordered and potentially glassy phase\cite{shinaoka2014effect,bilitewski2017jammed,dey2020random}. This can also be observed in the results from the low temperature snapshots and structure factor shown in Fig.~\ref{fig:figtodo}, where the disorder tends to blur the Bragg peaks. Notably, at higher values of the disorder amplitude ($\delta>1$), where for certain bonds the exchange coupling may shift from ferromagnetic to antiferromagnetic, and thus local frustration emerges, the chirality decreases, and the resulting textures become more disordered, as illustrated in Fig.~\ref{fig:figtodo}.

At higher fields, $B>0.6$, in Model I disorder reduces the chirality abruptly. In the skyrmion gas phase, even small amplitudes of disorder reduce the number of skyrmions, which get suppressed as $\delta$ is increased. In the field polarized phase, although strong disorder lowers the magnetization, chiral textures are not stabilized, and the chirality is practically zero at low temperatures, as seen in Fig.~\ref{fig:DP_last_T_bothmodels} (a).

In contrast,  Model II exhibits a particularly different response to disorder at higher fields. Here, we find that the chiral region expands with $\delta$, occupying a bigger region of the phase diagram across a range of magnetic fields and temperatures.  This is clearly seen comparing the evolution of the $B-T$ phase diagrams for both models, contrasting Fig.~\ref{DP_6_dJ} and Fig.~\ref{DP_6_dJ_M1}. Interestingly, at $B\sim0.6$ the skyrmion gas phase holds up to larger values of the disorder amplitude, which is reflected in the low temperature $B-\delta$ phase diagram presented in Fig.~\ref{fig:DP_last_T_bothmodels} (b). Moreover, in the field polarized region, a skyrmion gas emerges from the field polarized background at intermediate disorders ($\delta\sim 1$), and the system retains a non-zero chirality even as $\delta$ is further increased. 

The distinct effect of disorder and the emergent magnetic textures between both models is due to the way the disorder is introduced in each case. In Model I, fixing the ratio between the exchange and DM interactions in each bond is equivalent to proposing a random field model in each bond. Thus, at higher disorders, there is a significant probability of having bonds where the interactions are equivalent to having a higher magnetic field, and thus the topological textures are suppressed in the skyrmion gas region, which limits with the field polarized phase. On the other hand, in Model II the effect of disorder is random in each type of interaction, there is a competition between models with different $D_{ij}/J_{ij}$ ratios and therefore between different scales of skyrmion size and magnetic field. Thus, the topological skyrmion phases hold with larger disorder, and a skyrmion gas-like texture is induced by disorder from the field polarized phase. 

Our work highlights the importance of the interplay between magnetic frustration by competing interactions and disorder. By extensive Monte Carlo simulations and the implementation of a simple numerical model for disorder, we were able to gain valuable insights into the intricate behavior of magnetic systems under varying disorder conditions. This may set the stage for future investigation, experimentally or numerically, in the engineering of bulk systems with precisely defined sizes and density ranges. In this regard, our findings could be relevant for layered helimagnets with bond disorder, such as Mn$_{1-x}$Fe$_x$Ge \cite{grigoriev2013chiral}.
We hope our study will inspire further research into various types of quenched disorder in layered chiral magnets with tetragonal lattices, where the DM interaction acts only between nearest neighbors. These concepts could also prompt investigations into the stabilization or emergence of skyrmion phases in antiferromagnetic systems to see if the effect of quenched disorder is amplified or reduced compared to ferromagnetic cases. 

\section*{Acknowledgments} 

The authors are partially supported by CONICET (PIP 2021-112200200101480CO),  SECyT UNLP PI+D  X893 and PICT-2020 - SERIEA-03205. F. A. G. A. is also supported by CONICET through PIBAA 28720210100698CO.

\appendix*


%

\end{document}